\documentclass[twocolumn,a4paper,showpacs,showkeys,aps,nofootinbib,floatfix]{revtex4}

\usepackage{epsfig}
\usepackage{amsmath}
\usepackage{amssymb}
\usepackage{hhline}
\usepackage{subfigure}

\newcommand{\reaktion}{\mbox{$\vec{p}p\rightarrow\,pp\:\!\eta$ }}

\begin{document}
\title{Analysing power $\boldsymbol{A_y}$ in the reaction
$\boldsymbol{\vec{p}p\rightarrow\,pp\:\!\eta}$ close to threshold}
\author{P.~Winter$^{1}$,
H.-H.~Adam$^{2}$,
F.~Bauer$^{3}$,
A.~Budzanowski$^{4}$,
R.~Czy$\dot{\mbox{z}}$ykiewicz$^{5}$,
T.~G\"otz$^{1}$,
D.~Grzonka$^{1}$,
L.~Jarczyk$^{5}$,
A.~Khoukaz$^{2}$, 
K.~Kilian$^{1}$,
C.~Kolf$^{1}$,
P.~Kowina$^{1,6}$,
N.~Lang$^{2}$,
T.~Lister$^{2}$,
P.~Moskal$^{1,5}$,
W.~Oelert$^{1}$,
C.~Quentmeier$^{2}$,
T.~Ro$\dot{\mbox{z}}$ek$^{6}$,
R.~Santo$^{2}$,
G.~Schepers$^{1}$,
T.~Sefzick$^{1}$,
M.~Siemaszko$^{6}$, 
J.~Smyrski$^{5}$, 
S.~Steltenkamp$^{2}$,
A.~Strza{\l}kowski$^{5}$,
M.~Wolke$^{1}$,    
P.~W{\"u}stner$^{7}$,
W.~Zipper$^{6}$           
}
 
\affiliation{$^1$ Institut f{\"u}r Kernphysik, Forschungszentrum J\"{u}lich, D-52425 J\"ulich, Germany}
\affiliation{$^2$ Institut f{\"u}r Kernphysik, Westf{\"a}lische Wilhelms--Universit{\"a}t,  D - 48149 M{\"u}nster, Germany}
\affiliation{$^3$ I. Institut f\"ur Experimentalphysik, Universit\"at Hamburg, D-22761 Hamburg, Germany}
\affiliation{$^4$ Institute of Nuclear Physics, PL-31-342 Cracow, Poland}
\affiliation{$^5$ Institute of Physics, Jagellonian University, PL-30-059 Cracow, Poland}
\affiliation{$^6$ Institute of Physics, University of Silesia, PL-40-007 Katowice, Poland}
\affiliation{$^7$ Zentrallabor f{\"u}r Elektronik,  Forschungszentrum J\"{u}lich, D-52425 J\"ulich, Germany}
\date{\today}

\begin{abstract}
Measurements of the $\eta$ meson production with a polarised proton
beam in the reaction \reaktion have been carried out at an excess
energy of $Q= 40\,$MeV. The dependence of the analysing power $A_y$ on
the polar angle $\theta^*_q$ of the $\eta$ meson in the center of mass
system (CMS) has been studied. The data indicate the possibility of an
influence of p- and d-waves to the close to threshold $\eta$ production.
\end{abstract}

\maketitle

\section{Introduction\label{introduction}}
Several measurements on the $\eta$ meson production in the proton-proton 
interaction covering a 100\,MeV excess energy range were performed at 
different accelerators. The determined total cross sections~\cite{bergdolt:93, 
chiavassa:94, calen:96, calen:97, hibou:98, smyrski:00}, as well as their 
differential distributions~\cite{calen:98, tatischeff:00, moskal:01-2, abdelbary:02} 
triggered intensive theoretical investigations aiming to understand the production 
mechanism on the hadronic and quark-gluon level.

In the theoretical descriptions of the $\eta$-production in nucleon-nucleon 
collisions the excitation of the $S_{11}(1535)$ resonance plays a decisive role. 
The hitherto performed studies with the aim to describe the total cross section 
show a dominance of this virtual $S_{11}$ nucleon isobar in the close-to-threshold 
production of the $\eta$ meson. The excitation of this intermediate state results 
from a one meson exchange (e.g. $\pi,\ \eta$ or $\rho$) between the two nucleons 
followed by a strong coupling of the $\eta N$ system to the $S_{11}$. 

Near threshold the energy dependence of the total cross section results from a 
three-body phase space modified by a strong nucleon-nucleon final state 
interaction and a significant contribution of the attractive interaction in 
the $\eta p$ system. Since several existing models with different scenarios of the 
excitation describe the existing data well, a confrontation of the predicitions with 
other observables is needed in order to distinguish between them. The measurements with 
polarised beam should settle the on-going discussion whether the $\eta$ production is 
dominated by $\rho$~\cite{germond:90,laget:91,santra:98,gedalin:98}, $\omega$~\cite{vetter:91} 
or $\eta$~\cite{batinic:97} exchange. The interference between considered amplitudes causes a 
different behaviour -- depending on the assumed scenario -- e.g. of the $\eta$ meson angular 
distributions. These differences are too weak in the close-to-threshold region to discriminate 
between different models. Yet, the predictions of the analysing power depend crucially on the 
assumed mechanism~\cite{faeldt:01,nakayama:02}.

So far the only measurement of the analysing power has been performed~\cite{tatischeff:00} 
at an excess energy $Q=1805$\,MeV. In the present experiment, the analysing power close to 
the production threshold is determined and results for the interference terms from 
contributing partial waves are presented. A comparison with theoretical predictions 
will be discussed in section \ref{comparison}.

Section \ref{experiment} contains the description of the experiment and the method to 
extract the \reaktion events. The following section introduces definitions and gives a 
theoretical overview. In section \ref{results}, the results are presented. 

\section{Experiment\label{experiment}}
Measurements of the \reaktion reaction were performed at the internal experiment 
COSY-11~\cite{brauksiepe:96} at the COoler SYnchrotron COSY~\cite{maier:97nim} in J\"ulich 
with a beam momentum of p$_{beam}=2.096\,$GeV/c corresponding to an excess energy of $Q=40\,$MeV. 
During the experiment, cycles of about ten minutes for the two different beam polarisations 
were adjusted.  

\begin{figure}[ht]
\subfigure[\label{invmass}]{\epsfig{file=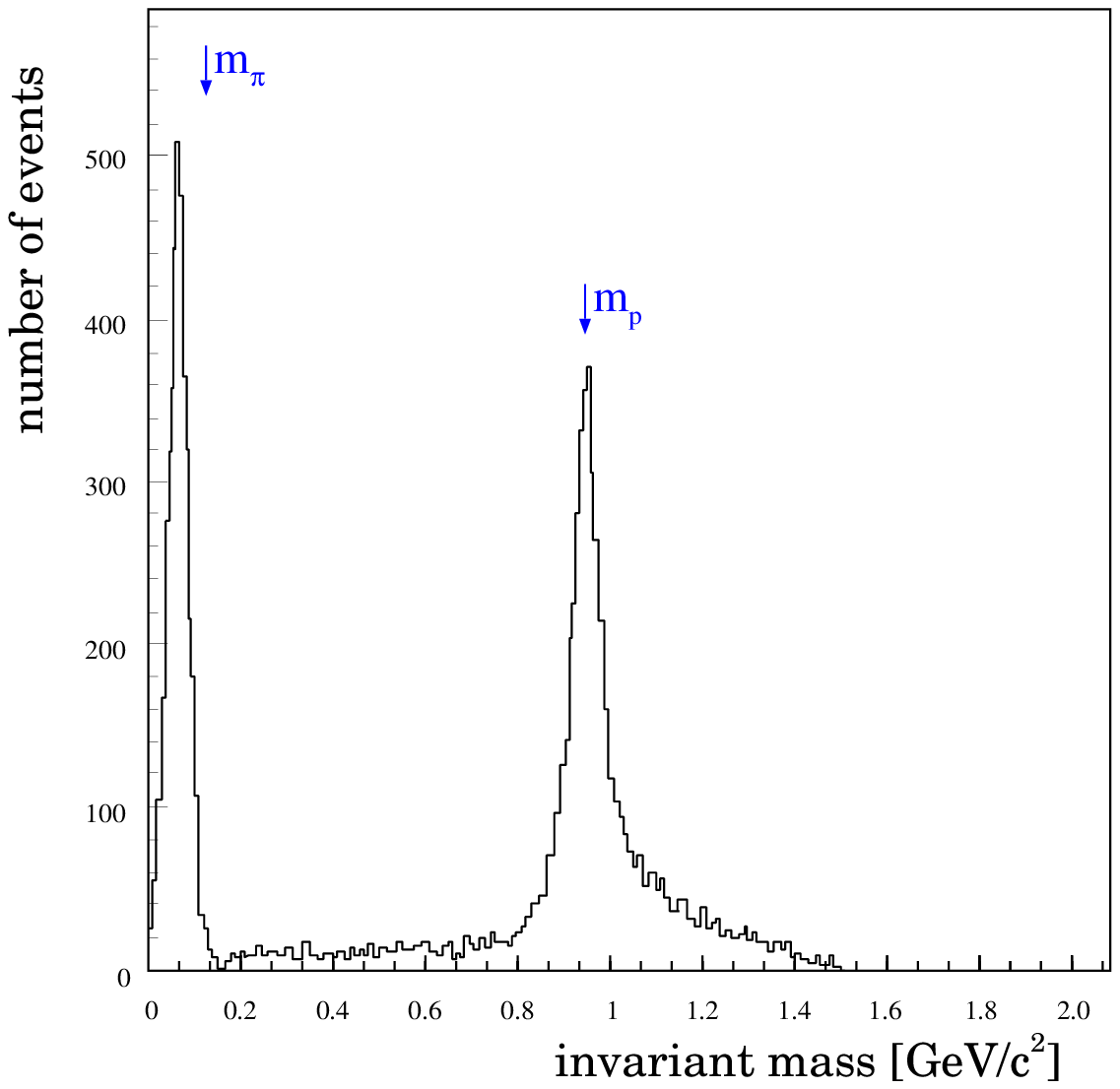,width=0.45\columnwidth}}
\subfigure[\label{missmass}]{\epsfig{file=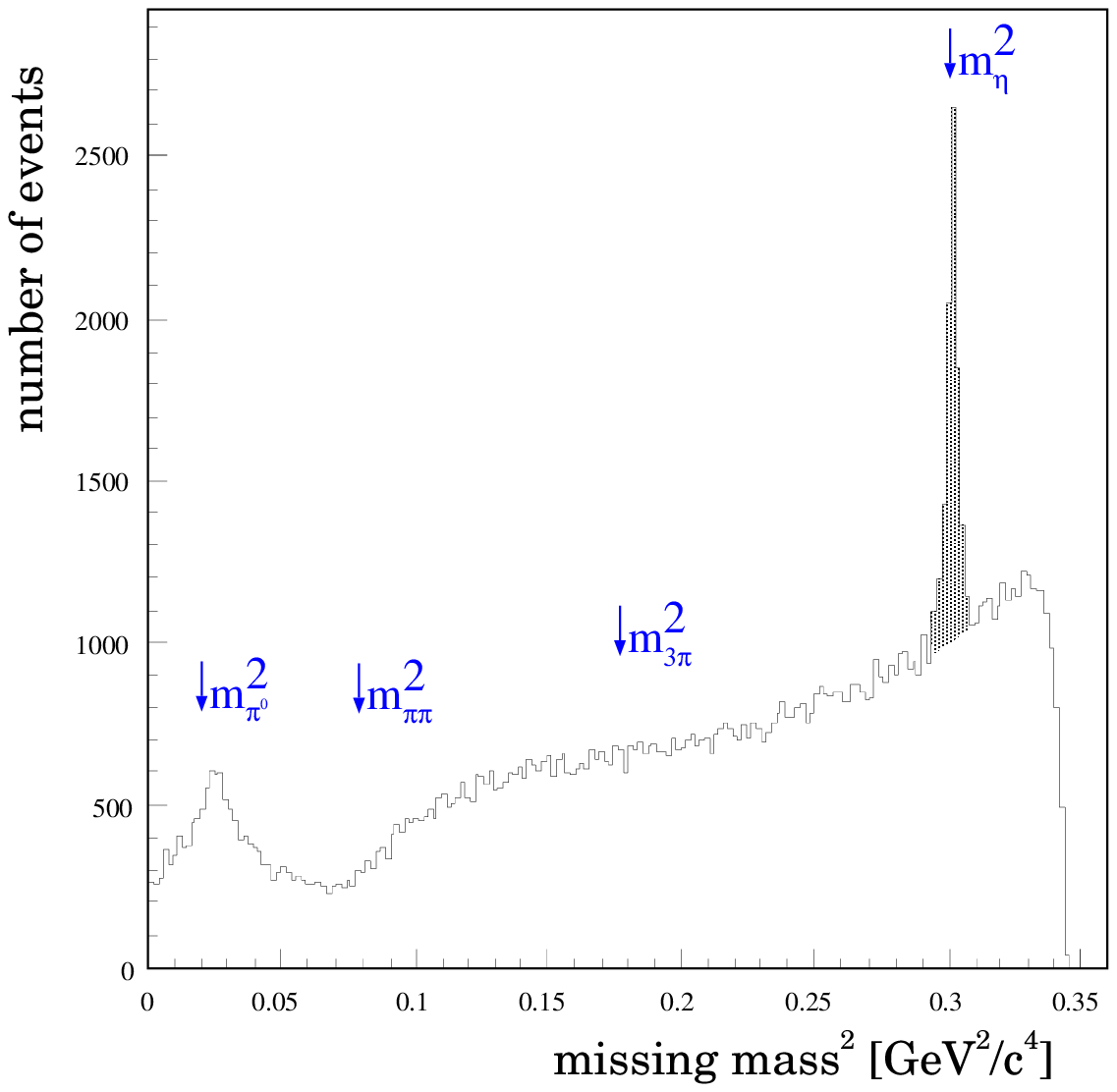,width=0.45\columnwidth}}
\caption{(a) Invariant mass spectrum for events with two reconstructed tracks. 
Besides the clear proton peak a second signal stemming from pions is observed. 
(b) Missing mass squared for events with two protons in the exit channel. 
Literature values~\cite{groom:00} for particle masses are indicated by arrows.}
\end{figure}

Using a hydrogen cluster target~\cite{dombrowski:97} in front of one of the regular COSY 
dipole magnets, the experimental facility acts like a magnetic spectrometer. Positively 
charged particles in the exit channel are bent towards the interior of the ring where 
they are detected in a set of two drift chambers. Tracing back the reconstructed trajectories 
through the magnetic field to the interaction point allows for momentum determination. 
Particle identification is achieved by a time of flight measurement over a distance of 9.4\,m 
between two scintillation hodoscopes. For further details, the reader is referred to 
reference~\cite{brauksiepe:96}. Figure \ref{invmass} shows that the method allows for a 
clear separation between pions and protons and hence for the identification of events 
with two protons in the exit channel.\\
The not registered particle system X -- either a single meson 
(the $\eta$ in the present case) or a multi meson system -- is identified by means 
of calculating its mass 
$m^2_X = (\mathbb{P}_{beam}+\mathbb{P}_{target} - \mathbb{P}_1 - \mathbb{P}_2)^2,$ 
while $\mathbb{P}_{beam}$ and $\mathbb{P}_{target}$ denote the four momentum of the 
beam and target proton in the initial channel and $\mathbb{P}_1,\ \mathbb{P}_2$ 
those of the two registered protons. The missing mass spectrum for events with two 
identified protons is shown in Figure \ref{missmass} for the entire beam time. 
Besides the clear $\eta$-signal there is obviously a $\pi^0$-peak resulting from 
the reaction \mbox{$\vec{p}p\rightarrow\,pp\:\!\pi^0$}. Furthermore, a broad yield 
due to multi pion events with the lower limit given by $m_X^2 = (2m_\pi)^2$ and the 
upper limit by $m_X^2 = (\sqrt{s}-2m_p)^2=0.345\,$GeV$^2$/c$^4$ is observed. The 
increasing event rate towards higher missing masses is due to the higher acceptance 
of the COSY-11 detector for two protons with small momenta in the center of mass 
system (CMS).

The monitoring of the geometrical dimensions of the synchrotron beam and its 
position relative to the target~\cite{moskal:01} enable to achieve a mass resolution 
of $\sigma_{m_\eta} = 1.6\,$MeV/c$^2$. The much broader peak of the $\pi^0$ is due 
to the error propagation which worsens the mass resolution with increasing excess 
energy~\cite{smyrski:00}.

\section{General Description}

\subsection{Definitions\label{definitions}}
A detailed theoretical derivation of the analysing power was recently published for 
the case of the $\vec{p}\vec{p}\rightarrow\,pp\pi^0$ 
reaction~\cite{meyer:01,meyer:99,meyer:98}. For the $\eta$ production the 
description is analogue since in both measurements the initial channel is 
fixed to isospin I=1. Therefore, the different quantum numbers for $\pi^0$ 
(as a member of an isotriplet) and the isoscalar $\eta$ are irrelevant. 

In the given experimental situation a convenient choice of the three axis is:
\begin{equation}
\hat{z} = \frac{\vec{p}_{beam}}{|\vec{p}_{beam}|}, \quad 
\hat{y} = \frac{\vec{P}}{|\vec{P}|} \quad\mbox{and}\quad
\hat{x} = \frac{\vec{y}\,\times\,\vec{z}}{|\vec{y}\,\times\,\vec{z}|} \label{achsen},
\end{equation}
where $\vec{P}$ indicates the polarisation of the COSY beam.

\begin{figure}[ht]
\begin{center}
\epsfig{file=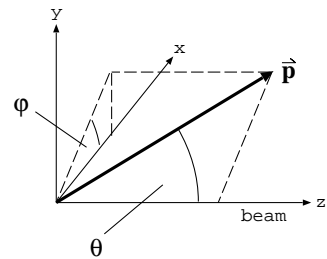,scale=0.7}
\end{center}
\caption{Definition of the angles. $\theta$ is defined as the angle 
between momentum vector and the z-axis, $\varphi$ between the x-axis and 
the projection of $\vec{p}$\, onto the x-y-plane.\label{winkel}}
\end{figure}

In the COSY-11 experiment, the two four momenta of the final protons 
$\mathbb{P}_i = (E_i^*, \vec{p}_i^{\,*})$ are measured. The CMS momentum of 
the $\eta$ meson is $\vec{q} = -(\vec{p}_1^{\,*}+\vec{p}_2^{\,*})$. The proton 
momentum in the pp rest-system is denoted by 
$\vec{p}$
For later purposes, Figure \ref{winkel} depicts the definition of the used polar- $(\theta)$ and 
azimuthal angle $(\varphi)$. The indices $p$ and $q$ will refer to the pp rest-system and the $\eta$ 
meson in the CMS, respectively. The angle $\theta_p$ will be choosen such, that 
$0\leq\theta_p\leq\pi/2$. This choice guarantees that all observables are 
invariant under the transformation $\vec{p} \rightarrow -\vec{p}$ as required 
by the identity of the two protons in the final state.

\subsection{Observables}
The differential cross section for a reaction with a polarised beam is given in 
terms of Cartesian polarisation observables by~\cite{fick:71}
\begin{equation}
\sigma(\xi)=\sigma_0(\xi)\left(1+\sum_{i=1}^3 P_i \cdot A_i(\xi) \right),\label{diffWQ}
\end{equation}
where $P_i$ and $A_i$ denote the beam polarisation and the analysing power in 
the given reference frame, $\sigma_0(\xi)$ indicates the total cross section in 
case of no polarisation. In the upper formula, the abbreviation 
\[
\sigma(\xi) = \frac{d^3 \sigma}{d\Omega_p d\Omega_q dE_{pp}}(\xi)
\]
is used where $\xi$ denotes the set of the five variables which are kinematically 
completely describing the exit channel, namely 
$(\theta_p,\ \varphi_p,\ \theta^*_q,\ \varphi^*_q,\ E_{pp})$. 
The kinetic energy $E_{pp}$ of the two final protons in their CM system is 
given by $E_{pp} = \sqrt{s_{12}} - 2m_p$ with 
$ \sqrt{s_{12}} = 2\sqrt{\vec{p}^2 + m_p^2}$ as the energy in the pp subsystem. 

In the given case of the general experimental conditions, the beam polarisation 
is -- due to the magnetic fields in the accelerator -- forced to be 
$\vec{P}=(0, P_y, 0)^T$ and hence formula \eqref{diffWQ} simplifies to
\begin{equation}
\sigma(\xi)=\sigma_0(\xi)\big(1+ P_y \cdot A_y(\xi) \big). \label{WQ}
\end{equation}

The asymmetry $\varepsilon$ -- obtained from the difference in the yields 
with beam polarisation up and down -- defined by
\begin{equation}
\varepsilon := \frac{N_\uparrow - N_\downarrow}{N_\uparrow + N_\downarrow} \label{epsilon}
\end{equation}
forms the basis for deducing the analysing power while $N_\uparrow\ (N_\downarrow)$ 
denote the experimental number of events for spin up (down). With known 
luminosity~$\mathcal{L}$, efficiency $\mathcal{E}$ and measured time $dt$, $dN$ is 
related to the cross section by 
$dN_{\uparrow,\downarrow} = \mathcal{E}\cdot\mathcal{L}\cdot \sigma_{\uparrow,\downarrow}\cdot dt$. 
In combination with equation \eqref{WQ}, one can deduce from \eqref{epsilon} that
\begin{equation}
 A_y(\xi) = \frac{\mathcal{L}_{rel}\cdot N_\uparrow - N_\downarrow}{
 N_\downarrow\cdot P_\uparrow - \mathcal{L}_{rel}\cdot P_\downarrow\cdot N_\uparrow}(\xi), 
 \label{formelfueray}
\end{equation}
where the relative time-integrated luminosity $\mathcal{L}_{rel}$ is defined by 
$\mathcal{L}_{rel}:=\frac{\int \mathcal{L}_\downarrow\cdot dt_\downarrow}
{\int \mathcal{L}_\uparrow\cdot dt_\uparrow}$. In equation \eqref{formelfueray}, 
the efficiency cancels out because of the independence on the spin as long as the 
bin size of $\Delta\xi$ is small enough so that the efficiency can be assumed to 
be constant.

With the definitions given in section \ref{definitions}, the angular dependence of 
the spin-dependent cross section can be written as~\cite{meyer:01}:
\begin{widetext}
\begin{eqnarray}
\sigma_0(\xi)A_y(\xi) & = &\Big\{\big[G_1^{y0} + G_2^{y0}
(3\cos^2\theta_p-1)\big]\sin\theta^*_q +\Big.\big[H_1^{y0}+I^{y0}+H_2^{y0}
(3\cos^2\theta_p-1)\big]\sin 2\theta^*_q\Big\}\cos\varphi^*_q \nonumber \\
& & \ \:\!{}+\Big.\big[H_3^{y0} + K^{y0} + G_3^{y0}\cos\theta^*_q + H_4^{y0}
(3\cos^2\theta^*_q-1)\big]\sin2\theta_p\cos\varphi_p \nonumber \\
& & {}+\Big.(G_4^{y0}\sin\theta^*_q + H_5^{y0}\sin 2\theta^*_q)\sin^2\theta_p\cos(2\varphi_p-\varphi^*_q)
 + H_6^{y0}\sin 2\theta_p\sin^2\theta^*_q\cos(2\varphi^*_q-\varphi_p) \label{spinWQ}
\end{eqnarray}
\end{widetext}
The appearing literals\footnote{The superscript $y0$ indicates a beam polarisation along 
the y-axis and an unpolarised target.} denote interferences of partial wave amplitudes. 
The relative angular momentum of the two outgoing protons in their rest system is denoted 
by capital letters $l_p=S,P,D\ldots$, the one of the $\eta$ meson in the CMS by small 
letters $l_q=s,p,d\ldots$, while the usual spectroscopic notation is used. 
With this definition, the single terms $G_k^{y0}, H_k^{y0}, I^{y0}$ and $K^{y0}$ 
correspond to (PsPp), (Pp)$^2$, (SsSd) and (SsDs).

\section{Results\label{results}}
In order to extract the assymetry from the measured spinup and spindown events one 
needs the relative luminosity and the average beam polarisation.

Via a simultaneous measurement of the proton-proton elastic scattering at the 
internal experiment EDDA~\cite{albers:97, altmeier:00} the polarisation was 
determined for two time blocks\footnote{The significant increase of the 
polarisation from the first to the second block is caused by improved tuning of 
the beam with respect to polarisation.}:\\[0.2cm]
\centerline{\begin{tabular}{|l|c||c|}
\hline
& time block 1 & time block 2 \\
\hhline{|==::=|}
P$_\uparrow$ & $0.381\pm 0.007$ & $0.497\pm 0.006$ \\
\hhline{|--||-|}
P$_\downarrow$ & $-0.498\pm 0.007$ & $-0.572\pm 0.007$ \\
\hline
\end{tabular}}
\\[0.1cm]

The relative luminosity $\mathcal{L}_{rel}= \frac{N^{elas}_\downarrow}
{\sigma^{elas}_\downarrow}\cdot \frac{\sigma^{elas}_\uparrow}{N^{elas}_\uparrow}$ 
was extracted via the elastic proton-proton scattering. To determine the elastic 
cross $\sigma^{elas}_{\uparrow,\downarrow}$ section according to equation \eqref{WQ} 
the analysing power was taken from~\cite{altmeier:00}. 
With the number of events $N^{elas}_{\uparrow,\downarrow}$ resulting from 
the elastic pp-scattering $\mathcal{L}_{rel}$ was calculated according to 
the definition given above:

\begin{center}
\begin{tabular}{|l|c||c|}
\hline
& time block1 & time block 2 \\
\hhline{|==::=|}
$\mathcal{L}_{rel}$ & $1.004\pm 0.004 {+ 0.002 \atop - 0.002}$ & 
$0.949 \pm 0.004 {+ 0.001 \atop - 0.001}$ \\
\hline
\end{tabular}
\end{center}

\begin{figure}[ht]
\begin{center}
\epsfig{file=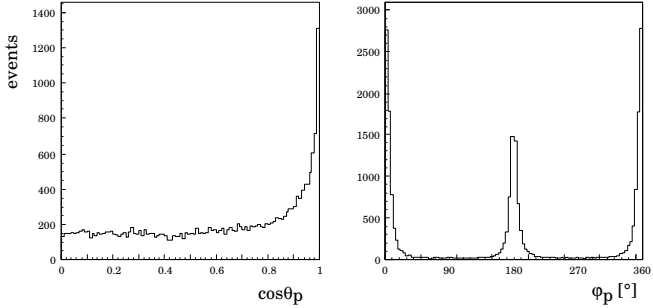, scale=0.5}
\end{center}
\caption{Angular distribution for the two proton angles $\cos\theta_p$ 
and $\varphi_p$ obtained via Monte-Carlo simulations.\label{efficiency}}
\end{figure}

An integration of equation \eqref{spinWQ} over $\cos\theta_p$ and $\varphi_p$ 
leads to the disappearance of several terms provided the experimental angular 
distribution covers either the full phase space with a constant detector 
efficiency or symmetrical ranges. Figure \ref{efficiency} shows the angular 
distributions of $pp\eta$-events from a Monte-Carlo simulation which are 
neither symmetric around 90$^\circ$ in case of $\varphi_p$ nor constant for 
both angles $\cos\theta_p$ and $\varphi_p$. Therefore, the evaluation of the 
analysing power requires an efficiency correction. To correct the data the 
efficiency $\mathcal{E}(\cos\theta_p,\varphi_p)$ is determined via Monte-Carlo 
simulations. Using a GEANT-3 code for each event a detection system response was 
calculated and the simulated data sample was analysed with the same programme 
which is used for the analysis of the experimental data. 
Weights $w(\cos\theta_p,\varphi_p)=1/\mathcal{E}(\cos\theta_p,\varphi_p)$ 
were applied during the final analysis of the experimental data and hence 
the corrected number of events reads:
\begin{equation}
N^{cor} = \frac{\sum\limits_{i}\sum\limits_{k} w_{i,k} N_{i,k}}
{\sum\limits_{i}\sum\limits_{k} w_{i,k}}\, , \label{efficiencycorrect}
\end{equation}
while $i$ and $k$ run over the bins $\varphi_p$ and $\cos\theta_p$, respectively. 
The error is deduced with $\Delta N_{i,k} = \sqrt{N_{i,k}}$ to be
\begin{equation*}
\Delta N^{cor} =  \frac{1}{\sum\limits_{i}\sum\limits_{k} w_{i,k}}
 \sqrt{\sum\limits_{i}\sum\limits_{k} w^2_{i,k} N_{i,k}}\,,
\end{equation*}
whereas the error of $w_{i,k}$ was neglected because of a much 
higher statistic for the Monte Carlo simulations, so that 
$\frac{\Delta w_{i,k}}{w_{i,k}} \ll\frac{\Delta N_{i,k}}{N_{i,k}}$. 
The influence of the strong proton-proton final state-interaction (FSI) 
was included via the description with a Jost-function~\cite{goldberger:64}. 
Former acceptance studies on the dependence on the various Jost function 
prescriptions showed a change of the result of maximum 10\%~\cite{moskal:00,moskal:00-2}. 
An extensive discussion on the influence of the FSI reflecting itself in the density
distribution of the Dalitz plot is given in~\cite{moskal_wolke}. 
Concerning that the FSI is known up to an accuracy of around 30\% one can conclude 
that the upper limit for the total contribution to the error is approximately 3\% 
which is negligible compared to the high overall error of this first data sample.

Only after an efficiency correction one can remove the dependency from 
proton-coordinates in the analysis which is then the same as an integration 
over these variables so that equation \eqref{spinWQ} simplifies to:
\begin{eqnarray}
\lefteqn{\iint \frac{d^2\sigma}{d\Omega_p d\Omega_q}(\xi)\,A_y(\xi)\, d\cos\theta_p\,d\varphi_p =} \nonumber \\
& & 2\pi\left(G_1^{y0}\sin\theta^*_q + (H_1^{y0}+I^{y0})\sin 2\theta^*_q\right) \cos\varphi^*_q.
\label{WQintegriert}
\end{eqnarray}
Due to the restricting dipole gap $\varphi^*_q$ is dominantly peaked around 0$^\circ$ -- quite 
similar to the $\varphi_p$ distribution -- but with a negligible peak around 180$^\circ$ 
which is not shown here but verified with MC simulations. Therefore, the analysis was 
performed with one single $\varphi^*_q$-bin around $\pm 30^\circ$. Hence, equation \eqref{WQintegriert} 
leads further to the separation of the (PpPs)-interference ($G_1^{y0}$) and the (Pp)$^2$- and 
(SsSd)-terms ($H_1^{y0}$ and $I^{y0}$):
\begin{eqnarray}
G_1^{y0} & = & \frac{1}{\pi^2} \int f(\cos\theta^*_q)\,d\cos\theta^*_q\, , \nonumber\\
\label{partialwellen} \\
H_1^{y0}+I^{y0} & = & \frac{2}{\pi^2}
 \int f(\cos\theta^*_q) \cos\theta^*_q \,d\cos\theta^*_q\, , \nonumber
\end{eqnarray}
with $f(\cos\theta^*_q)=\int\limits_{-\frac{\pi}
{6}}^{\frac{\pi}{6}}\int\limits_{0}^{2\pi}\int\limits_{0}^{1}
\frac{d^2\sigma}{d\Omega_p d\Omega_q}(\xi)\,A_y(\xi)\, 
d\cos\theta_p d\varphi_p d\varphi^*_q$.

Defining\footnote{In the following, limits of the integrations 
will be omitted as they are always the same.}  
$\bar{N}(\cos\theta^*_q) := \iiint N^{cor.}(\xi)\,d\Omega_p\,d\varphi^*_q$, 
it is straightforward to show analogue to equations \eqref{epsilon} and 
\eqref{formelfueray} that the integrated analysing power defined by
\begin{equation}
\bar{A}_y(\cos\theta^*_q):=f(\cos\theta^*_q)/\frac{d\sigma}{d\cos\theta^*_q} 
\label{integray}
\end{equation}
can be determined via
\begin{equation}
\bar{A}_y(\cos\theta^*_q) = 
\frac{\mathcal{L}_{rel}\cdot \bar{N}_{\eta,\uparrow} - \bar{N}_{\eta,\downarrow}}
{P_\uparrow\cdot \bar{N}_{\eta,\downarrow} - \mathcal{L}_{rel}\cdot P_\downarrow\cdot 
\bar{N}_{\eta,\uparrow}}(\cos\theta^*_q). \label{ayfuerauswertung}
\end{equation}

\begin{figure}[ht]
\begin{center}
\subfigure[\label{etamissmassmitu}]{\epsfig{file=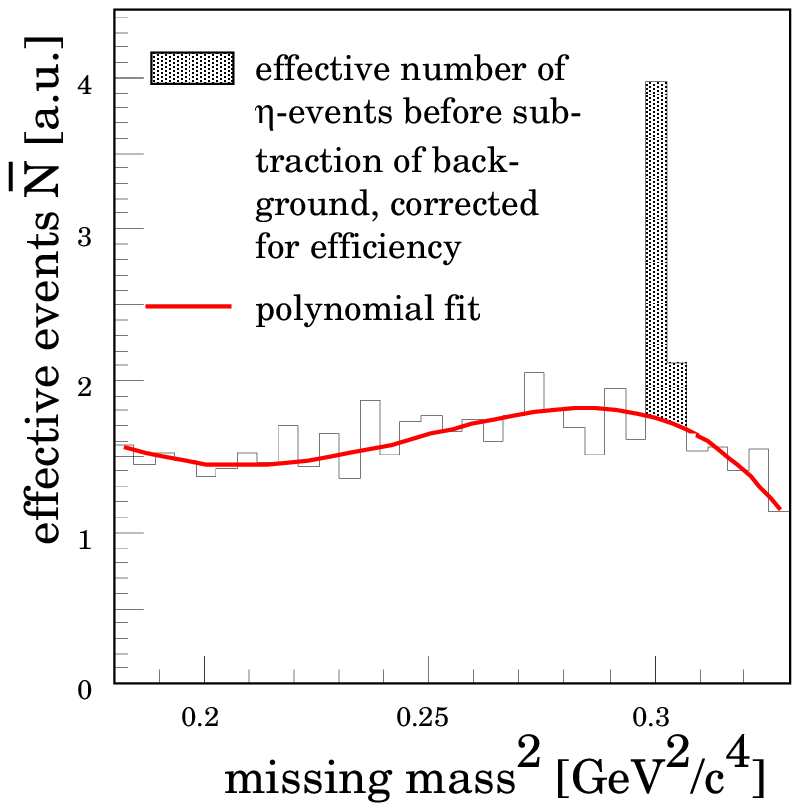,scale=0.5}}
\subfigure[\label{etamissmassohneu}]{\epsfig{file=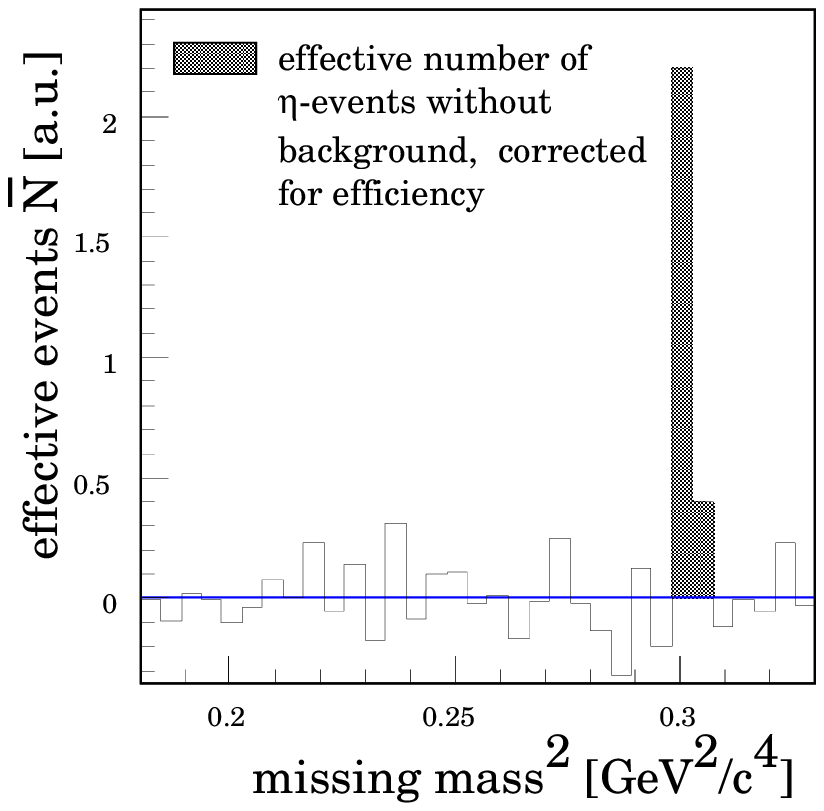,scale=0.5}}
\end{center}
\caption{Spectra of the squared missing mass for events with two identified protons 
(a) with and (b) after subtraction of the background. The number of events was 
corrected according to equation \eqref{efficiencycorrect}.}
\end{figure}

The calculation of $\bar{A}_y$ needs the determination of the absolute \reaktion 
events $\bar{N}_{\eta,\uparrow\downarrow}$ in dependence of $\cos\theta^*_q$. 
In section \ref{experiment}, the selection of the pp$\eta$ was discussed. The 
analysis was performed with 4 bins in $\cos\theta^*_q$ starting at 
$\cos\theta^*_q = -0.75$ with $\Delta\cos\theta^*_q = 0.5$. A representative 
missing mass spectrum is shown in Figure \ref{etamissmassmitu} where the 
background is fitted by a polynomial function. From this spectrum the number 
of events $\bar{N}_{\eta+b}$ including background and $\eta$-event are extracted. 
Subsequently, this background is subtracted and the number of events 
$\bar{N}_{\eta}$ are determined (Figure \ref{etamissmassohneu}).

\begin{figure}[htb]
\epsfig{file=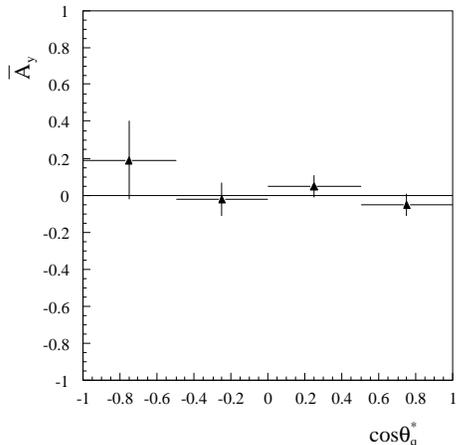,width=0.7\columnwidth}
\caption{Dependence of the analysing power on the center of mass polar angle 
$\theta^*_q$ of the $\eta$ meson. \label{ay-cos}}
\end{figure}

For the two time blocks, an error weighted mean value for $\bar{A}_y$ is calculated. 
Figure \ref{ay-cos} shows the analysing power as a function of $\cos\theta^*_q$. 
The extraction of $G_1^{y0}$ and $H_1^{y0}+I^{y0}$ with equations \eqref{partialwellen} 
needs according to \eqref{integray} the knowledge of $\frac{d\sigma}{d\cos\theta^*_q}$ 
which was taken from~\cite{calen:98}. The fact that 
$\varphi_q \in [-\frac{\pi}{6},\frac{\pi}{6}]$ was considered 
with $\frac{d\sigma}{d\cos\theta^*_q}=\int^{-\frac{\pi}{6}}_{\frac{\pi}{6}} 
\frac{d\sigma}{d\Omega^*_q}\,d\varphi^*_q$ which is due to the isotropy of the 
cross section in $\varphi$
\begin{equation}
\frac{d\sigma}{d\cos\theta^*_q}\bigg|\raisebox{-0.3cm}{$\scriptstyle -\frac{\pi}{6} 
\leq \varphi^*_q \leq \frac{\pi}{6}$} = \frac{\pi}{3} \frac{d\sigma}{d\Omega^*_q}. \label{diffwq}
\end{equation}

The averaged values of $\bar{A}_y$ and the cross section used for the integrations in 
equations \eqref{partialwellen} are presented in table~\ref{ayaveraged}.
\begin{table}[ht]
\begin{center}
\begin{tabular}{||c|c|c||}
\hhline{|t:===:t|} 
$\boldsymbol{\rule[-9pt]{0cm}{22pt}\cos\theta_q}$ & $\boldsymbol{\bar{A}_y}$ & $\boldsymbol{\frac{d\sigma}{d\cos\theta_q}}$ [$\mu$b] \\
\hhline{|:===:|}
-0.75\,$\pm$\,0.25 & 0.19\,$\pm$\,0.21 & 0.31\,$\pm$\,0.01 \\
\hhline{||---||}
-0.25\,$\pm$\,0.25 & -0.02\,$\pm$\,0.09 & 0.50\,$\pm$\,0.01 \\
\hhline{||---||}
 0.25\,$\pm$\,0.25 & 0.05\,$\pm$\,0.06 & 0.50\,$\pm$\,0.01 \\
\hhline{||---||}
 0.75\,$\pm$\,0.25 & -0.05\,$\pm$\,0.06 & 0.31\,$\pm$\,0.01 \\
\hhline{|b:===:b|}
\end{tabular}
\end{center}
\caption{Analysing power as a function of the emission angle $\theta_q$ of the $\eta$ 
meson in the CMS and the differential cross section obtained from~\cite{calen:98} with 
equation \eqref{diffwq}.\label{ayaveraged}}
\end{table}

Finally, the integrations of these values
\begin{eqnarray*}
G_1^{y0} & = & \frac{1}{\pi^2} \sum_{\cos\theta^*_q} 
\frac{d\sigma}{d\cos\theta^*_q} \bar{A}_y\cdot \Delta\cos\theta^*_q\\
H_1^{y0}+I^{y0} & = & \frac{2}{\pi^2} \sum_{\cos\theta^*_q} 
\frac{d\sigma}{d\cos\theta^*_q} \bar{A}_y \cos\theta^*_q\cdot \Delta\cos\theta^*_q\,.
\end{eqnarray*}
result in 
\[
G_1^{y0}= (0.003\pm 0.004)\,\mu\mbox{b}
\]
and 
\begin{equation*}
H_1^{y0}+I^{y0}=(-0.005\pm 0.005)\,\mu\mbox{b}\,. 
\end{equation*}

\section{Comparison with theory\label{comparison}}
The present data on the $\eta$ meson production in nucleon-nucleon collisions 
referred to in section \ref{introduction} show not only the 3-body phase space 
$Q^2$-dependency and a modification due to the nucleon-nucleon final state 
interaction but also a significant influence of the nucleon-meson interaction 
in the case of the $\eta p$ system. As mentioned above, several models describe 
the existing data quite well although they are based on different assumptions for 
the excitation mechanism of the $S_{11}$ resonance. For instance, 
Batini\'c et al.~\cite{batinic:97} or Nakayama, Speth and Lee~\cite{nakayama:02} 
found a dominance of $\pi$ and $\eta$-exchange in the analysis of 
$pp\rightarrow\,pp\:\!\eta$ while F\"aldt and Wilkin~\cite{faeldt:01} conclude a 
dominant $\rho$-exchange.

\begin{figure}[ht]
\begin{center}
\epsfig{file=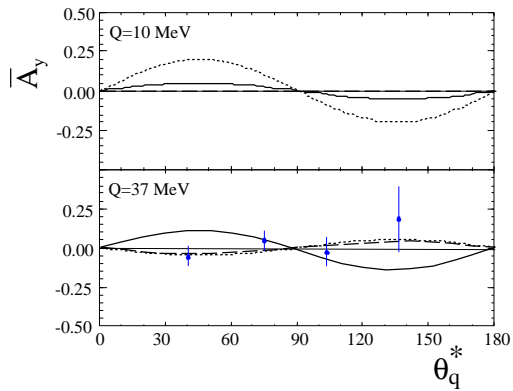, scale=0.85}
\end{center}
\caption{Analysing power for the reaction \reaktion in dependence on $\theta^*_q$ for 
the two excess energies $Q=10\,$MeV and 37\,MeV. \label{aypredict}}
\end{figure}

Polarisation observables may be the right tool to distinguish between the 
different models. Calculations for the analysing power in the reaction \reaktion show 
different results depending on the underlying assumption for the one meson exchange model. 
Figure \ref{aypredict} presents results taken from references~\cite{faeldt:01} (dotted line) 
and~\cite{nakayama:02} (solid and dashed lines) for $Q=10\,$MeV and 37\,MeV. The authors of 
the latter reference conclude in the full model calculations a dominance of $\pi$ and 
$\eta$-exchange (solid line). The dashed curve represents a vector dominance model 
with an exclusion of $\pi$ and $\eta$-exchange for exciting the $S_{11}$ resonance. 
The triangles are the experimental results. 

The observable structure of the experimental values show a slight deviation from 
the $\sin\theta_q\cos\theta_q$-dependence of both models. It seems that the data 
favours the vector dominance exchange models. The more or less strong difference 
in the angular dependency of $A_y$ results from a vanishing $G_1^{y0}$ in both 
references. As this corresponds to the (PpPs)-term, an influence of the P-wave 
must be suspected but right now the experimental result for $G_1^{y0}$ is compatible 
with zero. A non-zero $G_1^{y0}$ would imply that $H_1^{y0}$ -- describing the 
(Pp)$^2$ interference -- should have a non negligible contribution, too. For further 
detailed studies the data are not yet precise enough to disentangle the sum of 
$H_1^{y0}$ and $I^{y0}$. At this time the results indicate the possibility of an 
influence of p- and d-waves to the close to threshold $\eta$ production.

\section{Conclusion}
The reaction \reaktion has been studied at an excess energy of $Q=40\,$MeV. The 
final state has been kinematically completely reconstructed and the analysing 
power has been determined. Qualitatively, the data seem to favour the calculations 
with dominant vector meson exchange but definitive conclusions cannot be drawn due 
to the large uncertainties of the data. To allow a more rigorous comparison with 
theoretical calculations higher statistics experiments are required and already 
scheduled for 2002 at COSY-11.

\section{Acknowledgements}
We are very grateful for the support of the EDDA-collaboration in determining the 
beam polarisation. Furthermore, we would like to thank C. Wilkin, C. Hanhart and 
K. Nakayama for very helpful discussions and contributions. Our special thanks 
go to Prof. Dr. J. Treusch, chairman of the board of directors of the research 
center J\"ulich, for undertaking a nightshift during the experiment.

This work has been supported by the International B{\"uro} and the Verbundforschung 
of the BMBF, the Polish State Committee for Scientific Research,  the FFE grants 
from the Forschungszentrum J{\"u}lich, the Forschungszentrum J{\"u}lich 
directorates and the European Community - Access to Research Infrastructure 
action of the Improving Human Potential Programme.

\end{document}